%%%%%%%%%%%% Begin Macro %%%%%%%%%%%%%%%%%%%%%%%%%%%%%%%%%%%%%%%%%%%%%%%%%%%%%%%%%%%%%%%%%%%%%%

\catcode`@=11

\newskip\ttglue

\font\twelverm=cmr12 \font\twelvebf=cmbx12
\font\twelveit=cmti12 \font\twelvesl=cmsl12

\font\ninerm=cmr9
\font\eightrm=cmr8
\font\sixrm=cmr6
\font\eighti=cmmi8   \skewchar\eighti='177
\font\sixi=cmmi6     \skewchar\sixi='177
\font\ninesy=cmsy9   \skewchar\ninesy='60
\font\eightsy=cmsy8  \skewchar\eightsy='60
\font\sixsy=cmsy6    \skewchar\sixsy='60
\font\eightbf=cmbx8
\font\sixbf=cmbx6
\font\eighttt=cmtt8  \hyphenchar\eighttt=-1
\font\eightit=cmti8
\font\eightsl=cmsl8

\def\smalltype{\def\rm{\fam0\eightrm}
 			\textfont0=\eightrm  \scriptfont0=\sixrm  \scriptscriptfont0=\fiverm
 			\textfont1=\eighti   \scriptfont1=\sixi   \scriptscriptfont1=\fivei
 			\textfont2=\eightsy  \scriptfont2=\sixsy  \scriptscriptfont2=\fivesy
 			\textfont3=\tenex    \scriptfont3=\tenex  \scriptscriptfont3=\tenex
    \textfont\itfam=\eightit  \def\it{\fam\itfam\eightit}
	   \textfont\slfam=\eightsl  \def\sl{\fam\slfam\eightsl}
	   \textfont\ttfam=\eighttt  \def\tt{\fam\ttfam\eighttt}
    \textfont\bffam=\eightbf  \scriptfont\bffam=\sixbf
        \scriptscriptfont\bffam=\fivebf  \def\bf{\fam\bffam\eightbf}
    \tt  \ttglue=.5em plus.25em minus.15em
    \normalbaselineskip=9pt
    \setbox\strutbox=\hbox{\vrule height7pt depth2pt width0pt}
    \let\sc=\sixrm  \let\big=\eightbig  \normalbaselines\rm}
\def\eightbig#1{{\hbox{$\textfont0=\ninerm\textfont2=\ninesy
    \left#1\vbox to6.5pt{}\right.\n@space$}}}

\def\medtype{\def\rm{\fam0\tenrm}
 			\textfont0=\tenrm  \scriptfont0=\sevenrm  \scriptscriptfont0=\fiverm
 			\textfont1=\teni   \scriptfont1=\seveni   \scriptscriptfont1=\fivei
 			\textfont2=\tensy  \scriptfont2=\sevensy  \scriptscriptfont2=\fivesy
 			\textfont3=\tenex    \scriptfont3=\tenex  \scriptscriptfont3=\tenex
    \textfont\itfam=\tenit  \def\it{\fam\itfam\tenit}
	   \textfont\slfam=\tensl  \def\sl{\fam\slfam\tensl}
	   \textfont\ttfam=\tentt  \def\tt{\fam\ttfam\tentt}
    \textfont\bffam=\tenbf  \scriptfont\bffam=\sevenbf
        \scriptscriptfont\bffam=\fivebf  \def\bf{\fam\bffam\tenbf}
    \tt  \ttglue=.5em plus.25em minus.15em
    \normalbaselineskip=12pt
    \setbox\strutbox=\hbox{\vrule height8.5pt depth3.5pt width0pt}
    \let\sc=\eightrm  \let\big=\tenbig  \normalbaselines\rm}

\def\bigtype{\let\rm=\twelverm \let\bf=\twelvebf
\let\it=\twelveit \let\sl=\twelvesl \rm}

\def\footnote#1{\edef\@sf{\spacefactor\the\spacefactor}#1\@sf
    \insert\footins\bgroup\smalltype
    \interlinepenalty100 \let\par=\endgraf
    \leftskip=0pt  \rightskip=0pt
    \splittopskip=10pt plus 1pt minus 1pt \floatingpenalty=20000
  \vskip4pt\noindent\hskip20pt\llap{#1\enspace}
\bgroup\strut\aftergroup\@foot\let\next}
\skip\footins=12pt plus 2pt minus 4pt \dimen\footins=30pc

\def\bigfont{\magnification=1200 \baselineskip=20pt}

 \def\b{\beta} 
 \def\g{\gamma}

\def\cl#1{\centerline{#1}}
\def\clbf#1{\centerline{\bf #1}}

\def\is#1{{\narrower\smallskip\noindent#1\smallskip}}

\long\def\myname{\medskip
\cl{Kiho Yoon}
\cl{Department of Economics, Korea University}
\cl{145 Anam-ro, Seongbuk-gu, Seoul, Korea 02841}
\cl{ \tt kiho@korea.ac.kr}
\cl{\tt https://kihoyoon.github.io}
\medskip}

\def\ve{\vfill\eject}

\def\frac#1#2{{#1 \over #2}}
\def\Re{I\!\!R}

\newcount\sectnumber
\def\Section#1{\global\advance\sectnumber by 1 \bigskip
           \noindent{\bigtype {\bf \the\sectnumber  \ \ \ #1}} \medskip}

\def\prop#1{\medskip\noindent {\bf Proposition #1.} \it}
\def\lemma#1{\medskip\noindent {\bf Lemma #1.} \it}
\def\thm#1{\medskip\noindent {\bf Theorem #1.} \it}

\def\ok{\smallskip \rm}

\def\pf{\medskip\noindent Proof: \/}

\def\endpf{\hfill {\it Q.E.D.} \smallskip}

\def\app#1{\bigskip {\bigtype \clbf{Appendix #1}} \medskip}

\newcount\notenumber
\def\note#1{\global\advance\notenumber by 1
            \footnote{$^{\the\notenumber}$}{#1} \tenrm}

\def\ref{\bigskip \centerline{\bf REFERENCES} \medskip}

\def\emet{{\it Econometrica\/ }}
\def\jet{{\it Journal of Economic Theory\/ }}

\def\res{{\it Review of Economic Studies\/ }}
\def\geb{{\it Games and Economic Behavior\/ }}

\def\et{{\it Economic Theory\/ }}

\def\paper#1#2#3#4#5{\noindent\hangindent=20pt#1 (#2), ``#3,'' #4, #5.\par}

%%%%%%%%%%%% End Macro %%%%%%%%%%%%%%%%%%%%%%%%%%%%%%%%%%%%%%%%%%%%%%%%%%%%%%%%%%%%%%%%%%%%%%%%

\bigfont

\def\th{\theta}
\def\Th{\Theta}

{ \ }

\vskip 1cm

{\bigtype
\clbf{On the sufficiency of}
\clbf{unidirectional incentive compatibility in auctions}
}

\vskip 1cm
\bigskip

\myname

\vskip 0.5cm

\clbf{Abstract}
\is{\baselineskip=12pt We study optimal auction design when the direction of bidders' deviations is restricted. We show that the optimal revenue when bidders can only underbid their true values cannot exceed the optimal revenue when bidders may freely underbid or overbid. Thus, unidirectional incentive compatibility is sufficient for full incentive compatibility for revenue maximization. We prove this equivalence through linear programming duality in a discrete model, which makes it possible to analyze the feasibility of allocation rules in multi-agent environments.}

\smallskip

\is{\baselineskip=12pt Keywords: optimal auction, revenue maximization, incentive constraint, upper envelope, duality}
\smallskip

\is{\baselineskip=12pt  JEL Classification: C72; D44; D82}

\ve

\Section{Introduction}

This paper studies optimal auction design when the direction of bidders' deviations is restricted. The main result is that the optimal revenue when bidders can only underbid their true values cannot exceed the optimal revenue when bidders may freely underbid or overbid. Thus, unidirectional incentive compatibility is sufficient for full incentive compatibility for revenue maximization. We prove this equivalence through linear programming duality in a discrete model, which makes it possible to analyze the feasibility of allocation rules in multi-agent environments.

This paper can also be viewed as an alternative formulation of Myerson's (1981) optimal auction. We dispense with the monotonicity constraint on allocation rules, but instead incorporate the upper envelope formula into the objective function which accounts for the amount of information rent that must be conceded to bidders with higher values. Thus, a technical contribution of the present paper is that the upper envelope formulation replaces the monotonicity constraint.

This paper extends Moore's (1984) insight on the sufficiency of downward constraints to auctions with many bidders. He established this sufficiency for auctions with a single bidder who has finitely many possible values and who is risk averse, but left it an open question to solve for auctions with many bidders while pointing out the difficulty arising from the feasibility of allocation rules. The present paper is an answer to this question.

Celik (2006) and Kr\"ahmer and Strausz (2025) have inspired the present research. They have formulated unidirectional incentive compatibility and obtained several interesting results for the sufficiency in a principal-agent setting, the former with finite type spaces and the latter with continuous type spaces. Like Moore (1984), these works are also confined to the single-agent environments. To the best of our knowledge, the present paper is the first to show the sufficiency of downward incentive constraints for multi-agent environments.\note{Iyengar and Kumar (2008) and Malakhov and Vohra (2009) examine optimal auctions for capacity constrained bidders who can only underbid their capacities. These papers thus treat a special form of two-dimensional private information. They do not, however, discuss the sufficiency of unidirectional incentive compatibility.}

In the next section, we first characterize unidirectional incentive compatibility in terms of the upper envelope of the allocation rule and then formulate the revenue maximization problems. In section 3, we establish the sufficiency of downward constraints by investigating the dual programs. The analysis utilizes weak and strong duality, the lower-triangular decomposition of type matrices, and the concept of least concave majorant.\note{The analysis employing the least concave majorant corresponds to the ironing of virtual valuation in Myerson (1981).} In the final section, we briefly discuss how one can extend the result to continuous type spaces.

\Section{The formulation of auction problems}

There are $n$ bidders for an auction of one indivisible object. Each bidder has a privately-known value for the object, which is bidder $i$'s type. We assume that the set $\Th_i$ of bidder $i$'s types is finite so that $\Th_i = \{\th_{i1}, \ldots, \th_{iK_i}\}$ with $\th_{i1} < \cdots < \th_{iK_i}$. The probability of type $\th_{ir}$ is denoted by $p_{ir} > 0$, with $\sum_{r=1}^{K_i} p_{ir} = 1$. We assume that types are independent across bidders. We adopt the usual convention such as $\Th = \prod_{i=1}^{n} \Th_i$ and $\Th_{-i} = \prod_{j \ne i} \Th_j$. It is notationally convenient to denote a type profile $\th = (\th_{1k_1}, \cdots, \th_{ik_i}, \cdots, \th_{nk_n})$ by $k=(k_1,\cdots,k_i, \cdots, k_n) \in \prod_{i=1}^n\{1,\ldots,K_i\}$ and the probability of profile $k$ by $p(k) = \prod_{i=1}^n p_{ik_i}$. We also denote $p_{-i}(k_{-i}) = \prod_{j \ne i} p_{jk_j}$ and thus $p(k) = p_{ik_i}p_{-i}(k_{-i})$.

\medskip
\noindent {\sl 2.1. The mechanism}
\medskip
A direct mechanism is $(a, x)$ in which $a: \Th \rightarrow \Re^n_+$ is the allocation rule and $x: \Th \rightarrow \Re^n$ is the transfer rule.\note{It is without loss of generality to restrict our attention to direct mechanisms. For unidirectional incentive compatibility, especially, we can apply the nested range condition of Green and Laffont (1986). See also Kr\"ahmer and Strausz (2025).} Hence, $a_i(k)$ is the probability that bidder $i$ gets the object and $x_i(k)$ is the payment that bidder $i$ has to make when the type profile is $k$. The mechanism is {\it feasible\/} if
$$\sum_{i=1}^n a_i(k) \leq 1 \ \ \ \forall k. \eqno(0)$$
Define
$$A_{is} = \sum_{k_{-i}}p_{-i}(k_{-i})a_i(s,k_{-i}) {\rm \ \ and \ \ } X_{is} = \sum_{k_{-i}}p_{-i}(k_{-i})x_i(s,k_{-i}).$$
Thus, $A_{is}$ is the conditional expected allocation and $X_{is}$ is the conditional expected payment when bidder $i$ reports $\th_{is}$ to the mechanism. Bidder $i$'s expected payoff from reporting $\th_{is}$ when the true type is $\th_{ir}$ is defined as $\widetilde U_i(\th_{is};\th_{ir}) = A_{is} \th_{ir} - X_{is}$. We also define $U_{ir} = \widetilde  U_i(\th_{ir};\th_{ir})$. Hence,
$$U_{ir} = A_{ir} \th_{ir} - X_{ir}.$$

The mechanism $(a,x)$ is incentive compatible in the Bayesian sense if $\widetilde  U_i(\th_{ir};\th_{ir}) \geq \widetilde U_i(\th_{is};\th_{ir})$, that is, $U_{ir} \geq A_{is} \th_{ir} - X_{is}$ for all $i = 1, \ldots, n$ and $\th_{ir}, \th_{is} \in \Th_i$. Observe that the last inequality is equivalent to $U_{ir} \geq U_{is} + (\th_{ir} - \th_{is}) A_{is}$. We also consider the environments in which bidders can only underbid, but cannot overbid, their respective types. More precisely, the mechanism $(a,x)$ is unidirectionally incentive compatible (UIC) if
$$U_{ir} \geq U_{is} + (\th_{ir} - \th_{is}) A_{is} \qquad \forall i = 1, \ldots, n,  \forall s, r = 1, \ldots, K_i {\rm \ with \ } s \leq r. \eqno(1)$$
In comparison, the mechanism is (bidirectionally or fully) incentive compatible (IC) if
$$U_{ir} \geq U_{is} + (\th_{ir} - \th_{is}) A_{is} \qquad \forall i = 1, \ldots, n,  \forall s, r = 1, \ldots, K_i. \eqno(2)$$
The mechanism is (interim) individually rational (IR) if
$$U_{ir} \geq 0 \qquad \forall i = 1, \ldots, n,  \forall r = 1, \ldots, K_i. \eqno(3)$$

\medskip
\noindent {\sl 2.2. Unidirectional incentive compatibility}
\medskip

Let us define the {\it upper envelope\/} by
$$\bar A_{ist} = \max_{l = s, \ldots, t} A_{il}.$$
That is, $\bar A_{ist}$ is the smallest monotone upper envelope of $A_{il}$ for the region $s \leq l \leq t$. Observe that $\bar A_{ist}$ is increasing in $t = s, \ldots, K_i$.\note{We use the terms `increasing' and `decreasing' in the weak sense. We use `strictly increasing' and `strictly decreasing' for strong monotonicity.} Observe also that $\bar A_{ist} = A_{it}$ if $A_{it} = \max_{l=s, \ldots, t} A_{il}$ but $\bar A_{ist} > A_{it}$ otherwise. The following proposition characterizes unidirectional incentive compatibility.

\prop1  We have:
$$U_{ir} \geq U_{is} + (\th_{ir} - \th_{is}) A_{is} \qquad \forall i = 1, \ldots, n,  \forall s, r = 1, \ldots, K_i {\rm \ with \ } s \leq r \eqno(1)$$
if and only if
$$U_{ir} - U_{is} \geq \sum_{t=s}^{r-1} (\th_{i,t+1} - \th_{it}) \bar A_{ist}  \qquad \forall i = 1, \ldots, n,  \forall s, r = 1, \ldots, K_i {\rm \ with \ } s \leq r. \eqno(4)$$ \ok

\pf ($\Leftarrow$ part) Straightforward since
$$U_{ir} - U_{is} \geq \sum_{t=s}^{r-1} (\th_{i,t+1} - \th_{it}) \bar A_{ist} \geq \sum_{t=s}^{r-1} (\th_{i,t+1} - \th_{it}) A_{is} = (\th_{ir} - \th_{is}) A_{is},$$
where the second inequality follows from the definition of $\bar A_{ist}$.

($\Rightarrow$ part) We prove by induction on the number $r-s$. If $r-s=1$, then the right-hand side of (4) is equal to $(\th_{i, s+1} - \th_{is}) A_{is}$ since $\bar A_{iss} = A_{is}$. Thus, (4) is the same as (1). Next, assume the claim holds for natural numbers less than a given number $r-s$. Let $l^*$ be such that $l^* \in \arg\max_{l=s,\ldots,r-1} A_{il}$. We have $A_{il^*} = \max_{l=s,\ldots,r-1}A_{il}$ and thus $\bar A_{ist} = A_{il^*}$ for all $t = l^*, \ldots, r-1$. Now, by the induction hypothesis, we have
$$U_{il^*}-U_{is} \geq \sum_{t=s}^{l^*-1} (\th_{i,t+1}-\th_{it})\bar A_{ist}. \eqno(5)$$
Note that this inequality is valid even when $l^* = s$ since both sides become zero in that case. Next, $(1)$ gives $U_{ir} - U_{il^*} \geq (\th_{ir} - \th_{il^*}) A_{il^*}$. Since $\th_{ir} - \th_{il^*} = \sum_{t=l^*}^{r-1} (\th_{i,t+1} - \th_{it})$ and $\bar A_{ist} = A_{il^*}$ for all $t = l^*, \ldots, r-1$, we have
$$U_{ir} - U_{il^*} \geq \sum_{t=l^*}^{r-1} (\th_{i,t+1} - \th_{it}) \bar A_{ist}. \eqno(6)$$
By adding $(5)$ and $(6)$, we get $(4)$ and this completes the proof. \endpf

\medskip
\noindent {\sl 2.3. Revenue maximization}
\medskip

The seller's problem is to maximize the expected revenue
$$\sum_{i=1}^n \sum_{r=1}^{K_i} p_{ir} X_{ir}$$
subject to $(0), (1)$, and $(3)$ for unidirectional incentive compatibility, and subject to $(0), (2)$, and $(3)$ for (full) incentive compatibility. By the definition of the expected payoff, that is, by $U_{ir} = A_{ir} \th_{ir} - X_{ir}$, and Proposition 1, the seller's problem under unidirectional incentive compatibility is equivalent to the following:
$$\max_{a, U} \sum_{i=1}^n \sum_{r=1}^{K_i} p_{ir} (\th_{ir} A_{ir} - U_{ir})$$
subject to $(0)$, $(3)$, and $(4)$, where $U = (U_1, \cdots, U_n)$. With a slight abuse of notation, let
$$\bar A_{it} = \bar A_{i1t} = \max_{l = 1, \ldots, t} A_{il}.$$
That is, we drop the subscript $s$ from $\bar A_{ist}$ when $s=1$. Let us {\it relax\/} constraints $(3)$ and $(4)$ to
$$U_{i1} \geq 0 \qquad \forall i \eqno(7)$$
and
$$U_{ir} \geq U_{i1} + \sum_{t=1}^{r-1} (\th_{i, t+1} - \th_{it}) \bar A_{it} \qquad \forall i, \forall r. \eqno(8)$$
Indeed, $(3)$ implies $(7)$ and $(4)$ implies $(8)$. Since $U_{ir}$'s enter the objective negatively, both $(7)$ and $(8)$ must be binding at the solution. Therefore, we can replace inequalities with equalities in $(7)$ and $(8)$ and then merge these two constraints into the following single constraint.
$$U_{ir} = \sum_{t=1}^{r-1} (\th_{i, t+1} - \th_{it}) \bar A_{it} \qquad \forall i, \forall r. \eqno(9)$$
Thus, $(3)$ and $(4)$ imply $(9)$. On the other hand, observe that $(3)$ holds by $(9)$ and the fact $\bar A_{it} \geq 0$. Observe also that $(4)$ holds by $(9)$ since
$$U_{ir} - U_{is} = \sum_{t=s}^{r-1} (\th_{i, t+1} - \th_{it}) \bar A_{it} \geq \sum_{t=s}^{r-1} (\th_{i, t+1} - \th_{it}) \bar A_{ist},$$
where the equality holds by $(9)$ and the inequality holds since $\bar A_{ir} \geq \bar A_{isr}$ by definition.

By substituting $(9)$ into the objective of the seller's problem, we get
$$\sum_{i=1}^n \sum_{r=1}^{K_i} p_{ir} \th_{ir} A_{ir} - \sum_{i=1}^n \sum_{r=1}^{K_i} \sum_{t=1}^{r-1} p_{ir} (\th_{i, t+1} - \th_{it}) \bar A_{it}.$$
Since
$$\sum_{r=1}^{K_i} \sum_{t=1}^{r-1} p_{ir} (\th_{i, t+1} - \th_{it}) \bar A_{it} = \sum_{r=1}^{K_i -1} \Bigl[(\th_{i, r+1} - \th_{ir}) \sum_{t=r+1}^{K_i}p_{it}\Bigr] \bar A_{ir},$$
the revenue maximization problem under unidirectional incentive compatibility is
$$\max_{a: \Th \rightarrow \Re^n_+} \sum_{i=1}^n \sum_{r=1}^{K_i} p_{ir} \th_{ir} A_{ir} - \sum_{i=1}^n \sum_{r=1}^{K_i-1} c_{ir} \bar A_{ir} \leqno[S1]$$
subject to the feasibility constraint $(0)$, where we define
$$c_{ir} = (\th_{i, r+1} - \th_{ir}) \sum_{t=r+1}^{K_i} p_{it} \qquad \forall i, \forall r=1, \ldots, K_i-1. \eqno(10)$$
In comparison, the revenue maximization problem under (full) incentive compatibility constraint is well-known to be
$$ \max_{a: \Th \rightarrow \Re^n_+} \sum_{i=1}^n \sum_{r=1}^{K_i} p_{ir} \phi_{ir} A_{ir} \leqno[S2]$$
subject to the feasibility constraint $(0)$ and the monotonicity constraint
$$A_{ir} \leq A_{i, r+1} \qquad \forall i, \forall r = 1, \ldots, K_i - 1, \eqno(11)$$
where we define the (finite version of) {\it virtual valuation\/}
$$\phi_{ir} = \th_{ir} - \frac{c_{ir}}{p_{ir}}  \qquad \forall i, \forall r \eqno(12)$$
with the convention that $c_{iK_i}=0$.

The value of [S1] is greater than or equal to the value of [S2] since the constraint in the former is less stringent, that is, unidirectional incentive compatibility is implied by (full) incentive compatibility. It is straightforward to observe that they are the same when (i) the solution to [S1] is such that $A_{ir}$ is increasing in $r$ for all $i$, or (ii) the virtual valuation $\phi_{ir}$ is strictly increasing in $r$ for all $i$. When (i) holds, we have $\bar A_{ir} = A_{ir}$ and thus the objective of [S1] is the same as the objective in [S2]. When (ii) holds, the solution to [S2] is well-known to satisfy the monotonicity constraint $(11)$ and we again have $\bar A_{ir} = A_{ir}$. Do there exist cases when the value of [S1] is strictly greater than the value of [S2]? In the next section, we show that the value of [S1] is always equal to the value of [S2] by constructing and matching the dual problems.

\Section{Sufficiency of unidirectional incentive compatibility}

We first write the revenue maximization problems as linear programs. The linear programming formulation of the seller's problem under unidirectional incentive compatibility $(1)$ is
$$\max_{a:\Th \rightarrow \Re^n_+} \sum_{k} p(k)\sum_{i=1}^n \th_{ik_i}a_i(k) - \sum_{i=1}^n\sum_{r=1}^{K_i-1} c_{ir} \bar A_{ir} \leqno[P1]$$
subject to the feasibility constraint $(0)$ and the constraints that $A_{is} \leq \bar A_{ir}$ for all $i=1, \ldots, n$ and for all $s, r = 1, \ldots, K_i-1$ with $s \leq r$. Observe that [P1] is an ex-post version of [S1], recalling that $A_{is} = \sum_{k_{-i}}p_{-i}(k_{-i})a_i(s,k_{-i})$, $\bar A_{ir} = \max_{s = 1, \ldots, r} A_{is}$, and $c_{ir} = (\th_{i, r+1} - \th_{ir}) \sum_{t=r+1}^{K_i} p_{it}$. Similarly, the linear programming formulation of the seller's problem under (full) incentive compatibility constraint $(2)$ is
$$\max_{a:\Th \rightarrow \Re^n_+} \sum_{k} p(k)\sum_{i=1}^n \phi_{ik_i} a_i(k) \leqno[P2]$$
subject to the feasibility constraint $(0)$ and the monotonicity constraint defined in $(11)$ above. Observe that [P2] is an ex-post version of [S2], recalling that we define the virtual valuation $\phi_{ik_i}$ in $(12)$ above.

The dual of $[P1]$ has variables $\chi(k) \geq 0$ for all $k$ and $\mu_{irs} \geq 0$ for all $i=1, \ldots, n$ and for all $s, r = 1, \ldots, K_i-1$ with $s \leq r$. It is
$$\min \sum_{k} \chi(k) \leqno[D1]$$
subject to
$$\chi(k) + p_{-i}(k_{-i}) \sum_{r = k_i}^{K_i-1} \mu_{irk_i} \geq p(k)\th_{ik_i} \qquad \forall i, \forall k \eqno(13)$$
and
$$\sum_{s=1}^r \mu_{irs} \leq c_{ir} \qquad \forall i, \forall r=1, \ldots, K_i-1. \eqno(14)$$

The dual of $[P2]$ has variables $\chi(k) \geq 0$ for all $k$ and $\lambda_{ir} \geq 0$ for all $i=1, \ldots, n$ and for all $r=1, \ldots, K_i-1$ with boundary conventions $\lambda_{i0} = 0$ and $\lambda_{iK_i} = 0$. It is
$$\min \sum_{k} \chi(k) \leqno[D2]$$
subject to
$$\chi(k) + p_{-i}(k_{-i})(\lambda_{ik_i} - \lambda_{i, k_i-1}) \geq p(k) \phi_{ik_i} \qquad \forall i, \forall k. \eqno(15)$$
Thus, both $[D1]$ and $[D2]$ have the same objective function. Appendix A contains a detailed derivation of the dual programs.

Observe that, by the definition of the virtual valuation given in $(12)$, constraint $(15)$ can be written as
$$\chi(k) + p_{-i}(k_{-i})(c_{ik_i} + \lambda_{ik_i} - \lambda_{i, k_i-1}) \geq p(k) \th_{ik_i} \qquad \forall i, \forall k.$$
If we define $d_{ik_i} = c_{ik_i} + \lambda_{ik_i} - \lambda_{i,k_i-1}$, then this inequality becomes
$$\chi(k) + p_{-i}(k_{-i}) d_{ik_i} \geq p(k)\th_{ik_i}.$$
Comparing this inequality with $(13)$, we see that if we can choose $\mu_{irs}$ so that
$$\sum_{r = s}^{K_i-1} \mu_{irs} = d_{is} \eqno(16)$$
holds while respecting constraint $(14)$, then we can convert a feasible solution to $[D2]$ into a feasible solution to $[D1]$. The following proposition gives an affirmative answer.

\prop2 (Dual matching) Suppose a feasible $[D2]$ pair $(\chi,\lambda)$ satisfies, for each bidder $i$,
$$d_{ir}=c_{ir}+\lambda_{ir}-\lambda_{i,r-1} \geq 0 \quad \forall r \quad {\rm and} \quad \lambda_{i,K_i-1}=0. \eqno(17)$$
Then, there exists $\mu$ such that $(\chi,\mu)$ is feasible for $[D1]$. \ok

\pf Observe that $(17)$ implies
$$\sum_{s=1}^m d_{is} \geq \sum_{s=1}^m c_{is} \quad \forall m=1, \ldots, K_i-1 \quad {\rm and} \quad \sum_{s=1}^{K_i-1} d_{is} = \sum_{s=1}^{K_i-1} c_{is}$$
since $\lambda_{ir} \geq 0$ with $\lambda_{i0}=0$ as a convention. A bidder-by-bidder application of the lower-triangular decomposition given in Lemma 1 of Appendix B shows that there exist $\mu_{irs} \geq 0$ for $1 \leq s \leq r \leq K_i-1$ such that $(14)$ and $(16)$ hold.\note{In fact, the construction gives $(14)$ as an equality.} Therefore, $(\chi, \mu)$ is feasible for $[D1]$. \endpf

This proposition converts the monotonicity multipliers in $[D2]$ into the envelope multipliers in $[D1]$. Note that the objective value remains the same since it depends only on $\chi(k)$ and we transform $\lambda_{ir}$ to $\mu_{irs}$ while keeping $\chi(k)$ the same. Using this proposition, we establish that the seller's revenue under unidirectional incentive compatibility cannot exceed the seller's revenue under (full) incentive compatibility.

\thm1 The value of $[P1]$ is the same as the value of $[P2]$. \ok

\pf The value of $[P1]$ is greater than or equal to the value of $[P2]$ as discussed in the previous section. Thus, we need only to prove the reverse inequality. By strong duality for finite linear programs, the value of $[P2]$ is the minimum value of $[D2]$. We show in Lemma 2 of Appendix C that there exists an optimal solution $(\chi^*, \lambda^*)$ for $[D2]$ that satisfies
$$c_{ir}+\lambda^*_{ir}-\lambda^*_{i,r-1} \geq 0 \quad \forall r \quad {\rm and} \quad \lambda^*_{i,K_i-1}=0.$$
Proposition 2 in turn constructs $\mu^*$ such that $(\chi^*, \mu^*)$ is feasible for $[D1]$ with the same objective value $\sum_{k} \chi^*(k)$ as the optimal value of $[D2]$. Since $[D1]$ is a minimization problem, its optimal value is less than or equal to this value. We conclude that the value of $[P1]$ is less than or equal to the value of $[P2]$ by strong duality. In summary, we have the value of $[P1] = $ the value of $[D1] \leq \sum_{k} \chi^*(k) = $ the value of $[D2] = $ the value of $[P2]$. \endpf

\Section{Discussion}

We have established the result for finite type spaces. We now briefly discuss the extension to continuous type spaces. Assume that bidder $i$'s type $\th_i$ is drawn from $\Th_i = [\underline \th_i, \overline \th_i] \subseteq \Re_+$ according to a distribution $F_i$, with $f_i$ being the density. Define the conditional expected allocation $A_i(\th_i) = \int_{\Th_{-i}} a_i(\th_i, \th_{-i}) f_{-i}(\th_{-i}) d\th_{-i}$ and the upper envelope $\bar A_i(\th_i) = \sup_{\eta \in [\underline \th_i, \th_i]} A_i(\eta)$. The continuous analogue of $[P1]$ is
$$\max_{a: \Th \rightarrow \Re^n_+} \int_{\Th} \sum_{i=1}^n a_i(\th)\th_i f(\th) d\th - \sum_{i=1}^n \int_{\underline \th_i}^{\overline \th_i} \bar A_i(\th_i) (1-F_i(\th_i)) d\th_i$$
subject to the feasibility constraint that $\sum_{i=1}^n a_i(\th) \leq 1$ for all $\th$, and the continuous analogue of $[P2]$ is
$$\max_{a: \Th \rightarrow \Re^n_+} \int_{\Th} \Bigl[\sum_{i=1}^n \Bigl( \th_i - \frac{1-F_i(\th_i)}{f_i(\th_i)} \Bigr) a_i(\th) \Bigr] f(\th) d\th$$
subject to the feasibility constraint that $\sum_{i=1}^n a_i(\th) \leq 1$ for all $\th$ and the monotonicity constraint that $A_i(\th_i)$ is increasing in $\th_i$ for all $i$.

One route to the continuous case is to study a sequence of discrete environments with increasingly finer discretizations of $\Th_i$. This approximation approach is conceptually straightforward but is technically nontrivial: we need to carefully approximate $A_i(\th_i)$ and $\bar A_i(\th_i)$ as well as the continuous counterpart of the least concave majorant while preserving the feasibility condition. Another route is to directly prove the result by extending relevant concepts. This approach may pose more severe technical challenges. We encounter infinite-dimensional dual problems with nonnegative measures as multipliers. We leave both routes to future research.

\app{A}

In this appendix, we derive the two dual programs given in the text. For the primal problem $[P1]$, let $\chi(k) \geq 0$ for all $k$ be the dual variable corresponding to the feasibility constraint $\sum_{i=1}^n a_i(k) \leq 1$ of $(0)$ and $\mu_{irs} \geq 0$ for all $i=1, \ldots, n$ and for all $s, r = 1, \ldots, K_i-1$ with $s \leq r$ be the dual variable corresponding to constraint $A_{is} - \bar A_{ir} \leq 0$. Consider the primal variable $a_i(k)$ with $k_i = t$. It appears in the feasibility constraint for $k$ with coefficient 1 and in every constraint $A_{it} \leq \bar A_{ir}$ for $t \leq r$ with coefficient $p_{-i}(k_{-i})$. The primal objective coefficient is $p(k) \th_{it}$. This gives $(13)$ in $[D1]$. Next, consider the primal variable $\bar A_{ir}$. It appears in constraint $A_{is} \leq \bar A_{ir}$ for $s \leq r$ with coefficient $-1$. The primal objective coefficient is $-c_{ir}$. This gives $-\sum_{s=1}^r \mu_{irs} \geq - c_{ir}$, thus we have $(14)$ in $[D1]$. Since only the feasibility constraints have positive right-hand side of $1$, we get the dual objective of $\sum_{k} \chi(k)$.

For the primal problem $[P2]$, let $\chi(k) \geq 0$ for all $k$ be the dual variable corresponding to the feasibility constraint $\sum_{i=1}^n a_i(k) \leq 1$ of $(0)$ and $\lambda_{ir} \geq 0$ for all $i=1, \ldots, n$ and for $r=1, \ldots, K_i-1$ be the dual variable corresponding to the monotonicity constraint $A_{ir} - A_{i, r+1} \leq 0$ of $(11)$. Consider the primal variable $a_i(k)$ with $k_i = t$. It appears in the feasibility constraint for $k$ with coefficient $1$ and in the monotonicity constraint $A_{it} - A_{i, t+1}$ with coefficient $p_{-i}(k_{-i})$ and in the  monotonicity constraint $A_{i,t-1} - A_{it}$ with coefficient $-p_{-i}(k_{-i})$, recalling that $A_{it}=\sum_{k_{-i}} p_{-i}(k_{-i}) a_i(t, k_{-i})$. The primal objective coefficient is $p(k)\phi_{it}$. This gives $(15)$ in $[D2]$.  Since only the feasibility constraints have positive right-hand side of $1$, we get the dual objective of $\sum_{k} \chi(k)$.

\app{B}

We suppress the bidder index in the following lemma.

\lemma{1} (Lower-triangular decomposition) Let $K \geq 2$.  Suppose $c_r \geq 0$ for $r=1,\ldots,K-1$ and $d_s \geq 0$ for $s=1,\ldots,K-1$ satisfy
$$\sum_{s=1}^m d_s \geq \sum_{r=1}^m c_r \qquad \forall m=1,\ldots,K-1 \eqno(A1)$$
and
$$\sum_{s=1}^{K-1} d_s = \sum_{r=1}^{K-1} c_r. \eqno(A2)$$
Then, there exist $\mu_{rs} \geq 0$ for $1 \leq s \leq r \leq K-1$ such that
$$\sum_{s=1}^r \mu_{rs}=c_r \qquad \forall r=1, \ldots, K-1 \eqno(A3)$$
and
$$\sum_{r=s}^{K-1}\mu_{rs}=d_s \qquad \forall s=1, \ldots ,K-1. \eqno(A4)$$  \ok

\pf Let $M=K-1$. It is helpful to think of the subscript $r$ as an index for the rows and the subscript $s$ as an index for the columns for an $M \times M$ matrix. We prove the claim by induction on $M$.  The case $M=1$ is immediate from $(A2)$: set $\mu_{11}=c_1=d_1$. Next, assume the claim holds for $M-1$ columns and rows and consider a problem with $M$ rows and columns. By subtracting $(A1)$ for $m = M-1$ from $(A2)$, we get $c_M \geq d_M$. Set $\mu_{MM} = d_M$. The remaining mass in row $M$ is $c_M - d_M \geq 0$. Since row $M$ can put mass to all columns $1,\ldots,M-1$, we combine this residual mass with the mass of row $M-1$. Define a reduced $(M-1)$-row problem by
$$c'_r=c_r \quad {\rm for} \quad r=1, \ldots, M-2, \qquad c'_{M-1}=c_{M-1}+c_M-d_M$$
and
$$d'_s=d_s \quad {\rm for} \quad s=1,\ldots,M-1.$$
We have
$$\sum_{s=1}^{M-1}d'_s = \sum_{s=1}^{M-1}d_s = \sum_{r=1}^{M}c_r-d_M = \sum_{r=1}^{M-1}c'_r$$
as well as, for $m < M-1$,
$$\sum_{s=1}^m d'_s = \sum_{s=1}^m d_s \geq \sum_{r=1}^m c_r =\sum_{r=1}^m c'_r.$$
Hence, the induction hypothesis gives $\b_{rs} \geq 0$ for $1 \leq s \leq r \leq M-1$ satisfying the corresponding row sum $(A3)$ and the column sum $(A4)$ conditions for the reduced problem.

It remains only to split the last reduced row, whose mass is $c'_{M-1}=c_{M-1}+c_M-d_M$, into the original row $M-1$ with mass $c_{M-1}$ and the residual part of row $M$ with mass $c_M-d_M$.  Since both original rows $M-1$ and $M$ can put mass to every column $s \leq M-1$, this split is possible: choose nonnegative numbers $\g_s \leq \b_{M-1,s}$ with $\sum_{s=1}^{M-1}\g_s=c_{M-1}$, which exist because $\sum_s \b_{M-1,s} = c'_{M-1} \geq c_{M-1}$. Set
$$\mu_{M-1,s}=\g_s, \qquad \mu_{M,s}=\b_{M-1,s}-\g_s \quad {\rm for} \quad s=1, \ldots, M-1$$
and set $\mu_{rs} = \b_{rs}$ for $1 \leq s \leq r \leq M-2$. Recall that we set $\mu_{MM} = d_M$. The row sums are $c_1, \ldots, c_M$ and the column sums are $d_1, \ldots, d_M$. This completes the induction. \endpf

\app{C}

\lemma{2} (Finite ironing) There exist optimal $\lambda_{ir}$'s for $[D2]$ such that, for each bidder $i$,
$$c_{ir} + \lambda_{ir} -\lambda_{i,r-1} \geq 0 \quad \forall r \quad {\rm and} \quad \lambda_{i,K_i-1} = 0.$$ \ok

\pf Let $q_{ir} = \sum_{s=r}^{K_i} p_{is}$ with $q_{i,K_i+1}=0$ as a convention, and define the finite revenue points $R_{ir} = \th_{ir} q_{ir}$ with $R_{i,K_i+1} = 0$ as a convention. Consider the piecewise-linear function $R_i(q)$ that connects the points $(q_{ir}, R_{ir})$ for $r=1,\ldots, K_i+1$. Thus, $R_{ir} = R_i(q_{ir})$ and the slope of this function on the interval $[q_{i,r+1}, q_{ir}]$ is $(R_{ir}-R_{i,r+1})/(q_{ir}-q_{i,r+1})$. Since $q_{ir} - q_{i,r+1} = p_{ir}$, we have $R_{ir}-R_{i,r+1} = \th_{ir}(p_{ir}+q_{i,r+1}) - \th_{i,r+1} q_{i,r+1} = p_{ir} \th_{ir} - (\th_{i,r+1}-\th_{ir})q_{i,r+1} = p_{ir} \th_{ir} - c_{ir} = p_{ir} \phi_{ir}$, where the third equality follows from the definition $(10)$ of $c_{ir}$ and the last equality follows from the definition $(12)$ of virtual valuation $\phi_{ir}$. Thus,
$$p_{ir} \phi_{ir} = R_{ir} - R_{i,r+1} \eqno(A5)$$
which means that the slope of $R_i(q)$ on the interval $[q_{i,r+1}, q_{ir}]$ is equal to the virtual valuation $\phi_{ir}$.

Let $\overline R_i$ be the least concave majorant of $R_i$. That is, $\overline R_i$ is concave, satisfies $R_i(q) \leq \overline R_i(q)$ for every $q$, and is pointwise no greater than any other concave majorant of $R_i$.\note{That is, for any concave function $\hat R_i$ satisfying $R_i(q)\leq \hat R_i(q)$ for every $q$, we have $\overline R_i(q) \leq \hat R_i(q)$ for every $q$.} Let $\overline R_{ir} = \overline R_i(q_{ir})$. The slopes of $\overline R_i$ implicitly define the ironed virtual valuation $\overline \phi_{ir}$ by
$$p_{ir} \overline \phi_{ir} = \overline R_{ir} - \overline R_{i,r+1}. \eqno(A6)$$
Since $\overline R_i$ is concave in $q$, these slopes are increasing in the type index $r$.\note{Note well that $q_{ir}$ is decreasing in $r$.}

Consider an ironing block  $B=\{s,s+1,\ldots,t\}$, that is, a maximal consecutive set of types on which the least concave majorant $\overline R_i$ is linear. On this block, $\overline R_i$ is the chord from $(q_{is},R_{is})$ to $(q_{i,t+1},R_{i,t+1})$, and the common ironed virtual valuation is
$$\overline \phi_{iB} = \frac{R_{is} - R_{i,t+1}}{q_{is} - q_{i,t+1}}.$$
If $t < K_i$, then
$$\overline \phi_{iB}  = \frac{\th_{is} q_{is} - \th_{i,t+1} q_{i,t+1}}{q_{is}-q_{i,t+1}} = \th_{is}-\frac{(\th_{i,t+1}-\th_{is})q_{i,t+1}}{q_{is}-q_{i,t+1}} \leq \th_{is}.$$
If $t = K_i$, then $\overline \phi_{iB} = \th_{is} q_{is}/q_{is} = \th_{is}$ since $q_{i,K_i+1}=0$ and $R_{i,K_i+1}=0$. In either case, for every $r \in B$,
$$\overline \phi_{ir} \leq \th_{ir} \eqno(A7)$$
since $\th_{is} \leq \th_{ir}$.

Let us define
$$\lambda_{ir} = \overline R_{i,r+1} - R_{i,r+1} \eqno(A8)$$
for $r=0, 1, \ldots, K_i$. By the definition of the concave majorant, we have $\lambda_{ir} \geq 0$ for all $r$ with $\lambda_{i0}=\overline R_{i1} - R_{i1}=0$ and $\lambda_{iK_i} = \overline R_{i,K_i+1} - R_{i,K_i+1} = 0$. From $(A5)$, $(A6)$, and $(A8)$, we get $p_{ir} \overline \phi_{ir} = \overline R_{ir}-\overline R_{i,r+1} = (R_{ir}+\lambda_{i,r-1})-(R_{i,r+1}+\lambda_{ir}) = p_{ir} \phi_{ir} + \lambda_{i,r-1} - \lambda_{ir}$. That is,
$$p_{ir} \overline \phi_{ir} = p_{ir} \phi_{ir} - \lambda_{ir} + \lambda_{i,r-1}. \eqno(A9)$$

We show that $\lambda_{ir}$'s are optimal for $[D2]$. Multiplying both sides of $(A9)$ by $p_{-i}(k_{-i})$ and rearranging, we get $p(k) \overline \phi_{ik_i} + p_{-i}(k_{-i}) (\lambda_{ik_i}-\lambda_{i,k_i-1}) = p(k)\phi_{ik_i}$. By defining
$$\chi(k) = p(k) \max\{0, \max_{j} \overline \phi_{jk_j}\}, \eqno(A10)$$
we have $\chi(k) + p_{-i}(k_{-i}) (\lambda_{ik_i}-\lambda_{i,k_i-1}) \geq p(k)\phi_{ik_i}$ for all $i$ and $k$. Thus, $(\chi, \lambda)$ is feasible for $[D2]$. We next demonstrate optimality. Let $a^*: \Th \rightarrow \Re^n_+$ be a pointwise maximizer of
$$\sum_k p(k) \sum_{i=1}^n \overline \phi_{ik_i} a_i(k)$$
subject to the feasibility constraint $(0)$. Thus, $a^*_i(k) > 0$ only if $\overline \phi_{ik_i} = \max \{0, \max_j \overline \phi_{jk_j}\}$ for each $k$. Since $\overline \phi_{ir}$ is increasing in $r$ for all $i$, such a maximizer can be chosen in a way that the conditional expected allocation $A^*_i$ satisfies the monotonicity constraint $(11)$ for all $i$. Moreover, we can choose the tie-breaking so that adjacent types in the same ironing block receive the same allocation for each $k_{-i}$. Then, whenever $\lambda_{ir} > 0$, types $r$ and $r+1$ lie in the same ironing block and thus $A^*_{ir} = A^*_{i,r+1}$. This gives us $\lambda_{ir} (A^*_{ir} - A^*_{i,r+1}) = 0$ for all $i$ and $r$. Hence, we get
$$\sum_{r=1}^{K_i} p_{ir} \overline \phi_{ir} A^*_{ir} = \sum_{r=1}^{K_i} p_{ir} \phi_{ir} A^*_{ir} \eqno(A11)$$
by multiplying both sides of $(A9)$ by $A^*_{ir}$ and then summing over $r$ since\note{Recall that $\lambda_{i0}=\lambda_{iK_i}=0$.}
$$\sum_{r=1}^{K_i} (\lambda_{ir} - \lambda_{i, r-1}) A^*_{ir} = \sum_{r=1}^{K_i-1} \lambda_{ir}(A^*_{ir} -A^*_{i, r+1}) = 0.$$
Observe that
$$\eqalign{&\sum_k p(k) \sum_{i=1}^n \phi_{ik_i} a^*_i(k) = \sum_{i=1}^n \sum_{r=1}^{K_i} p_{ir} \phi_{ir} A^*_{ir} = \sum_{i=1}^n \sum_{r=1}^{K_i} p_{ir} \overline \phi_{ir} A^*_{ir} = \cr
= &\sum_k p(k)\sum_{i=1}^n \overline \phi_{ik_i} a^*_i(k) = \sum_{k} p(k) \max \{0, \max_j \overline \phi_{jk_j}\} = \sum_{k} \chi(k),}$$
where the first and third equalities hold by definition of conditional expected allocation, the second equality holds by $(A11)$, the fourth equality holds by definition of $a^*$, and the last equality holds by $(A10)$. The first term is the value of $[P2]$ at a feasible solution $a^*$ whereas the last term is the value of $[D2]$ at a feasible solution $(\chi, \lambda)$. We conclude by weak duality that both are optimal, in particular, $(\chi,\lambda)$ is optimal for $[D2]$.

Substituting $p_{ir} \phi_{ir} = p_{ir} \th_{ir} - c_{ir}$ into $(A9)$ gives
$$p_{ir}(\th_{ir} - \overline \phi_{ir}) = c_{ir} + \lambda_{ir} - \lambda_{i,r-1}. \eqno(A12)$$
By $(A7)$, the left-hand side is nonnegative. Therefore,
$$c_{ir} + \lambda_{ir} - \lambda_{i,r-1} \geq 0 \qquad \forall r.$$

It remains to show $\lambda_{i,K_i-1} = 0$. We first claim that the ironing block containing the highest type $K_i$ cannot contain any lower type. Suppose for the sake of contradiction that this block is $\{s, \ldots, K_i\}$ with $s < K_i$. Then, the chord from $(q_{is}, R_{is})$ to $(0,0)$ has value $\th_{is} q_{iK_i}$ at $q_{iK_i}$, which is strictly less than $R_{iK_i} = \th_{iK_i} q_{iK_i}$ because $\th_{is} < \th_{iK_i}$. Such a chord cannot majorize the revenue point $(q_{iK_i}, R_{iK_i})$, contradicting the definition of $\overline R_i$. Hence, the ironing block containing the highest type $K_i$ is the singleton $\{K_i\}$ and so $\overline \phi_{iK_i} = \th_{iK_i}$. We have, by $(A12)$ applied to $r = K_i$ together with the facts that $c_{iK_i}=0$ and $\lambda_{iK_i}=0$,
$$0 = p_{iK_i}(\th_{iK_i} - \overline \phi_{iK_i}) = c_{iK_i} +\lambda_{iK_i} - \lambda_{i,K_i-1} = -\lambda_{i,K_i-1}.$$
Thus, $\lambda_{i,K_i-1} = 0$ and this completes the proof. \endpf

\ref

\paper{Celik, G.}{2006}{Mechanism design with weaker incentive compatibility constraints}{\geb 56}{37-44}

\paper{Green, J., Laffont, J.-J.}{1986}{Partially verifiable information and mechanism design}{\res 53}{447-456}

\paper{Iyengar, G., Kumar, A.}{2008}{Optimal procurement mechanisms for divisible goods with capacitated suppliers}{{\it Review of Economic Design\/} 12}{129-154}

\paper{Kr\"ahmer, D., Strausz, R.}{2025}{Unidirectional incentive compatibility}{\jet 228}{106051}

\paper{Malakhov, A., Vohra, R.}{2009}{An optimal auction for capacity constrained bidders: a network perspective}{\et 39}{113-128}

\paper{Moore, J.}{1984}{Global incentive constraints in auction design}{\emet 52}{1523--1535}

\paper{Myerson, R.}{1981}{Optimal auction design}{{\it Mathematics of Operations Research\/} 6}{58-73}

\bye